\documentclass{elsarticle}

\usepackage[nolist]{acronym}
\usepackage{todonotes}
\usepackage{url}
\usepackage{graphicx}

\makeatletter
\def\ps@pprintTitle{%
 \let\@oddhead\@empty
 \let\@evenhead\@empty
 \def\@oddfoot{\centerline{\thepage}}%
 \let\@evenfoot\@oddfoot}
\makeatother

\begin{acronym}
\acro{TRA}{Technology Readiness Assessment}
\acro{ToA}{Target of Assessment}
\acro{TRL}{Technology readiness level}
\acro{STEM}{Software Technology Evaluation Model }
\acro{SMART}{Software Maturity Assessment and Readiness Technique }
\acro{ENISA}{European Union Agency for Cyber Security    }
\acro{CTE}{Critical Technology Elements }
\acro{COBIT}{Control Objectives for Information Technologies }
\acro{MRL}{Manufacturing Readiness Level }
\acro{SRL}{System Readiness Level    }
\acro{IRL}{Integrated Readiness Level  }
\acro{DoD}{Department of Defence }
\acro{DHS}{Department of Homeland Security }
\acro{DoE}{Department of Energy  }
\acro{TMP}{Technology Maturation plan }
\acro{NASA}{National Aeronautics and Space Agency }
\acro{CSC}{Critical Security Controls }
\acro{GDPR}{General Data Protection Regulation } 
\acro{NISD}{Directive on security of network and information systems}
\acro{TML}{Technology Maturity Level}
\end{acronym}

\begin{document}

\begin{frontmatter}

\title{SMART: a Technology Readiness Methodology in the Frame of the NIS Directive}
\tnotetext[t1]{The research for this article was funded by the
Luxembourg National Research Fund (FNR) C18/IS/12639666/En-
CaViBS/Cole, https://www.fnr.lu/projects/the-eu-nis-directive-enhancing-
cybersecurity-across-vital-business-sectors-encavibs/.}

\author{Archana Kumari$^1$, Stefan Schiffner$^2$, and Sandra Schmitz$^1$}
\address{$^1$University of Luxembourg, Esch-sur-Alzette, Luxembourg \\   
Sandra.Schmitz@uni.lu \\
$^2$ University of Münster, Münster, Germany \\
Stefan.Schiffner@uni-muenster.de}

\begin{abstract}
An ever shorter technology lifecycle engendered the need for assessing new technologies w.r.t. their market readiness. Knowing the  \ac{TRL} of a given target technology proved to be useful to mitigate risks such as cost overrun, product roll out delays, or early launch failures.\footnote{Steven Hirshorn and Sharon Jefferies, ``Final Report of the NASA Technology Readiness Assessment (TRA) Study Team" (2016).}
Originally developed for space programmes by NASA, \ac{TRL} became a de facto standard among technology and manufacturing companies and even among research funding agencies.\footnote{Mihály Héder, ``From NASA to EU: the evolution of the TRL scale in Public Sector Innovation" Innovation Journal (2017) Vol(22).}
However, while \ac{TRL} assessments provide a systematic evaluation process resulting in meaningful metric, they are one dimensional: they only answer the question if a technology can go into production. Hence they leave an inherent gap, i.e., if a technology fulfils requirements with a certain quality. This gap becomes intolerable when this metric is applied software such as technological  cybersecurity measures, cf Mankins.\footnote{John C Mankins, ``Technology Readiness Levels" white paper (1995)  \url{https://aiaa.kavi.com/apps/group\_public/download.php/2212/TRLs\_MankinsPaper\_1995.pdf} accessed 15 July 2021.} 

With legislation such as the General Data Protection Regulation\footnote{Regulation (EU) 2016/679 of the European Parliament and of the Council of 27 April 2016 on the protection of natural persons with regard to the processing of personal data and on the free movement of such data, and repealing Directive 95/46/EC (General Data Protection Regulation) [2016] OJ L119/1} (GDPR) and the Network and Information Systems Directive\footnote{Directive (EU) 2016/1148 of the European Parliament and of the Council of 6 July 2016 concerning measures for a high common level of security of network and information systems across the Union [2016] OJ L194/1.} (NIS-D) making reference to state of the art when requiring appropriate protection measures, software designers are faced with the question how to measure if a technology is suitable to use. We argue that there is a potential mismatch of legal aim and technological reality which not only leads to a risk of non-compliance, but also might lead to weaker protected systems than possible. In that regard, we aim to address the gaps identified with existing \ac{TRA}s  and aim to overcome these by developing standardised method which is suitable for assessing software w.r.t. its market readiness and quality (in sum maturity). 


\end{abstract}


\end{frontmatter}


\section{Introduction}

In the last decade, instances of cybercrime and -thefts have taken centre stage as the world has ushered into an internet-based economy. Some of the largest IT and software development firms have been the target of cyberattacks resulting in significant economic loss reaching billions of dollars.\footnote{Steve Morgan ``Cybercrime Facts and Statistics. 2021 Report: Cyberwarfare in the C-Suite" (Cybersecurity Ventures, 2021) \url{https://cybersecurityventures.com/wp-content/uploads/2021/01/Cyberwarfare-2021-Report.pdf} accessed June 2021.} 
For instance, according to the special report by Cyberwarfare in the C-Suite
, it is predicted to inflict damages amounting to 6 trillion USD world-wide in 2021; considering cybercrime as an economy, it would thus be the third largest economy in the world after the U.S. and China.\footnote{Ibid.}  Further, cybercrime, affects not only business entities or the state as such, also individuals are inter alia subject to blackmail and fraud, and their (physical) safety may be at risk. The world also saw the impact of abuse of data throughout the American presidential election in 2016, where data harvested by Cambridge Analytica was used to display customized messages about candidates to Facebook users in order to entice them to vote for their customer.
\footnote{Channel 4 News ``Cambridge Analytica CEO: We Ran all the Digital Campaign" (BBC News, 20 March 2018) \url{https://www.bbc.com/news/av/uk-43480751} accessed June 2021; for more analytical view on the matter J Isaak and M Hanna ``User Data Privacy: Facebook, Cambridge Analytica, and Privacy Protection" (2018) 8 Computer 56–59.}
  In response to an evolving cybersecurity threat landscape, the European Union is  taking political and legislative initiatives to make the Europe fit for the digital age. Legislative initiatives include the \ac{GDPR} and the \ac{NISD}, which both require service providers to implement \emph{state of the art} (technical and organisational) information security protection. Therefore, there is a need for a systematic process that integrates security and data protection requirements right from the early design process, including a way to assess the maturity/readiness of those characteristics at specified periods through a development lifecycle. This is important so that the burden of meeting these security requirements does not simply sit on companies but a solution on how to address/identify gaps are also developed that has standard understanding and acceptance in the software world.
\ac{TRA} methodologies have become increasingly popular across industries and businesses these days. \ac{TRA} is seen as a systematic, data driven, evidence-based process to evaluate the readiness of a technology,  the \ac{ToA} to be part of a larger system and fulfil ``mission critical" functions in said system. Here readiness means ready to use and cable to fulfil a given function. Up to today, \acp{TRL} are rate on a 9 step scale originally introduced by Mankins\footnote{John C Mankins (n3).} 
and revised by Hirshorn and Jefferies: each level is ordered sequentially as per the data evidence and performance/milestones achieved by the \ac{ToA}.\footnote{Steven Hirshorn and Sharon Jefferies (n 1).} 
Here the \ac{TRL} is a projection of the technology development process and depends on the data generated during the live cycle of the system. 

It is important to note that \ac{TRA} measures the usefulness of a \ac{ToA} for a specific environment and purpose. With the technological advancements and the inherent change of the environment that comes along with it, \ac{TRL} may not hold true for future purposes, however,  past evaluations do form a historical evidence on technology  progression.\footnote{Ibid.} 

While \ac{TRA} has proven effective in evaluating technology risk and identifying readiness level gaps for  \ac{ToA}, it has fallen short when applied to more service oriented \ac{ToA}. For example, \ac{TRA} provides readiness assessment snapshot for a specific point in time, that means the change in technology (and by that in environment) would make those evaluations stale and invalid. Therefore, the evaluation needs to be conducted and updated regularly, to keep up with the pace of technological advancement of the industry, with data evidences recorded for audit purposes, see also Hansen et al.\footnote{Marit Hansen and others, ‘Report on the workshop on assessing the maturity of privacy enhancing technologies in IFIP International Summer School on Privacy and Identity Management’ (2017)  1-15.}

Current \ac{TRA} methodologies do not clarify whether subsequent evaluation should be comprehensive or only focus on critical technology elements (CTEs) or how frequently audit should happen for CTEs.{\footnote{Nazanin Azizian, Thomas Mazzuchi and Sarkani Shahram, ``A comprehensive review and analysis of maturity assessment approaches for improved decision support to achieve efficient defense acquisition" (Proceedings of the World Congress on Engineering and Computer Science" 2009, Vol( 2).}\textsuperscript{,}\footnote{ ICEAA Professional Development and Training Workshop, ``TRL vs Percent Dev Cost" (2017) \url{http://iceaaonline.com/ready/wp-content/uploads/2017/07/DA05-PPT-Linick-Technology-Readiness-Level.pdf} accessed 15 July 2021.}} 
Further, \ac{TRA} does not measure gaps or risk for system cost, schedule or performance goals as part of technology risk evaluation.\footnote{United States Government Accountability Office,``Best Practices for Evaluating the Readiness of Technology for Use in Acquisition Programs and Projects" (2020) 99-102 \url{https://www.gao.gov/assets/710/703694.pdf} accessed 15 July 2021.} 

In addition, as per \ac{NASA} internal audit on \acp{TRA}/ \acp{TRL} confirms that TRL definitions are not specific or clear to the involved experts (program/product managers/developers), leaving assessment understanding on individual assessors, resulting in conflicting evaluations for a given \ac{ToA}.{\footnote{ibid 97.}\textsuperscript{,}\footnote{Steven Hirshorn and Sharon Jefferies (n 1).}\textsuperscript{,}\footnote{United States Government Accountability Office(n 16 ).} 


Beyond these general challenges of \ac{TRA}, specific challenges during the evaluation  of Software or Communication technologies have been identified.\footnote{Marit Hansen and others (n 13) 2-3.} 
Most of these issues emerged due to historical reasons since when the \acp{TRA} were initially designed, software technology was still in the early stages of evolution and hence software specific evaluation requirements were not considered for \acp{TRA}.\footnote{United States Government Accountability Office(n 16 ) 97.} However, the higher complexity of software products, i.e., a software can be in many more states than any mechanical device, opens a much larger space of possible solutions for a given functionality. In turn, not only not all solutions are equal, but also harder to compare w.r.t. the quality of the solution. So evaluating if a technology is ready to use alone is not sufficient anymore.

The key issue in applying existing \acp{TRA} to the software industry is due to the lack of metrics assessing qualities like cyber security and privacy aspects of the technology.\footnote{Marit Hansen and others, (n 13) 2-15.} 
Here three factors are immanent for cyber security: (1) quality aspects that describe none-functional requirements, (2) the product live cycle, and (3) the higher degree of composition of IT products, i.e. newly developed technology that needs to be integrated within a (potentially legacy) system.\footnote{ibid 2-5.}\textsuperscript{,}\footnote{ICEAA Professional Development and Training Workshop (n 15) 1-28.}
Hence, software developers have raised objection in usefulness of existing \acp{TRA} without customisation.\footnote{Steven Hirshorn and Sharon Jefferies (n 1) 20-27.}

To address the gaps identified and provide similar technology readiness evaluation standards as exists for other industries, attempts have been made with some success by  ISO\footnote{\url{www.itgovernance.co.uk}.}/ISA\footnote{\url{www.isa.org}.}/COBIT\footnote{\url{www.isaca.org/resources/cobit}.}/ CSC\footnote{\url{www.cisecurity.org/}.} and other such organizations. 

While NASA\footnote{\url{www.nasa.gov}.},  DoD\footnote{\url{www.acqnotes.com}.}, DoE, US GAO and other such organizations worldwide have evolved TRAs for engineering/manufacturing projects where the readiness demonstrations are physically visible, not much has happened on software industry front. 
Similar rules as in vanilla TRAs cannot be applied in case of software technology readiness evaluation.\footnote{Marit Hansen and others, (n 13) 2-15.}\textsuperscript{,}\footnote{Marit Hansen, Jaap-Henk Hoepman and Meiko Jensen, ‘Readiness Analysis for the Adoption and Evolution of Privacy Enhancing Technologies: Methodology, Pilot Assessment, and Continuity Plan’ (2015) Vol(1) 13-16.}

While those newly adopted evaluation methodes have addressed the concerns highlighted above to some extent, they have not been fully adopted by software community yet. Hence, there is critical need for a more unified TRA method for software technology maturity, i.e. combined quality and readiness,  evaluation to increase preparedness in the software industry sector against risks such as cyber attacks and data thefts.\footnote{ibid}
A detailed description of the gaps observed from these variations of TRLs for usage in software technology space is then detailed in subsequent sections. A novel way of TRL meant for software technology assessment is thus proposed drawing from those gaps analyzed.

We propose a new evaluation method: Software Maturity Assessment and Readiness Technique (SMART), which will showcase aspects of software and information technology that need to be considered in such an assessment, but have been under represented. Further, we supplement our maturity assessment method with a handbook for software \acp{TRL}. To achieve this goal, the paper will start with research findings on existing \acp{TRL} and will explain why those \acp{TRA} lack accurate maturity assessment capability for software technologies. 

Finally, the recommendation on the process and a sample Software TRL assessment calculator will be provided that will form part of Software \ac{TRL} handbook. A proposal on validation/beta testing of this approach, methodology and the handbook are also suggested and planned in the final section. This will help validate that the theory has real world application and would result in correct evaluation of any software technology readiness and maturity assessment.

\section{The notion of ‘state of the art’ from a legal perspective}

In the context of IT security, EU legal instruments commonly demand that IT security is brought up to the level of ‘state of the art’.

As the first horizontal instrument on cybersecurity at EU level, the NIS Directive.\footnote{Directive (EU) 2016/1148 of the European Parliament and of the Council of 6 July 2016 concerning measures for a high common level of security of network and information systems across the Union [2016] OJ L 194/1}  requires that Member States shall ensure that operators of essential services (OESs) and digital service providers (DSPs) take appropriate and proportionate technical and organisational measures to manage the risks posed to the security of network and information systems which they use in their operations, or as regards DSPs in the context of offering services referred to in Annex III of the Directive.\footnote{Arts. 14(1) and 16(1) NIS Directive.}  Having regard to the ‘state of the art’, those measures shall ensure a level of security of NIS appropriate to the risk posed.\footnote{Ibid.}   Similarly, Arts. 25 and 32 GDPR\footnote{Regulation (EU) 2016/679 of the European Parliament and of the Council of 27 April 2016 on the protection of natural persons with regard to the processing of personal data and on the free movement of such data, and repealing Directive 95/46/EC (General Data Protection Regulation) [2016] OJ L119/ 1.}  require data controllers, and to some extent processors, to take the ‘state of the art’ into account when implementing appropriate technical and organisational measures to mitigate the risks caused by their data processing activities. According to  Art. 40(1) EECC,\footnote{Directive (EU) 2018/1972 of the European Parliament and of the Council of 11 December 2018 establishing the European Electronic Communications Code [2018] OJ L 321/36 (EEEC).} The same applies to public electronic communications networks or services regarding the security of their networks and services.

None of these legal interventions provides a binding legal definition of the concept of ‘state of the art’ in the context of IT security. 

If one consults the vast body of EU legislation, the notion ‘state of the art’ is primarily used in legal acts relating to environment and technology such as for instance the Medical Devices Regulation\footnote{Regulation (EU) 2017/745 of the European Parliament and of the Council of 5 April 2017 on medical devices, amending Directive 2001/83/EC, Regulation (EC) No 178/2002 and Regulation (EC) No 1223/2009 and repealing Council Directives 90/385/EEC and 93/42/EEC, OJ L 117/1.}, the Radio Equipment Directive\footnote{Directive 2014/53/EU of theEuropean Parliament and of the Council of 16 April 2014 on the harmonisation of the laws of the Member States relating to the making available on the market of radio equipment and repealing Directive 1999/5/EC, OJ L 153/62.}, or the Machinery Directive.\footnote{Directive 2006/42/EC of the European Parliament and of the Council of 17 May 2006 on machinery, and amending Directive 95/16/EC, OJ L 157/24.} 

Although the notion is widely referred to in legal texts, there is no standard legal definition of the notion. Legal scholars thus refer to the notion as an indefinite, abstract general notion\footnote{See Martini in: Paal/Pauly, DS-GVO BDSG (3rd ed. 2021), Art. 25 DS-GVO marginal no. 39a.}, or undetermined legal concept\footnote{Von Grafenstein, Co-Regulation and the Competitive Advantage in the GDPR: Data protection certification mechanisms, codes of conduct and the ‘state of the art’ of data protection-by-design; in: González Fuster, van Brakel, de Hert, Research Handbook on Privacy and Data Protection Law: Values, Norms and Global Politics.}. 

This raises the question, what can be considered ‘state of the art’.

In the past EU legislation also referred in the context of technology to ‘best available techniques’ for which for instance Directive 2010/75/EU\footnote{Directive 2010/75/EU of the European Parliament and of the Council of 24 November 2010 on industrial emissions (integrated pollution prevention and control, OJ L 334/17.}  provides a definition in the context of emissions.  Art. 3(10) Directive 2010/75/EU defines ‘best available techniques’ in the context of emissions as ‘the most effective and advanced stage in the development of activities and their methods of operation which indicates the practical suitability of particular techniques for providing the basis for emission limit values and other permit conditions designed to prevent an, where that is not practicable, to reduce emissions and the impact on the environment as a whole: (a) ‘techniques’ includes both the technology used and the way in which the installation is designed, built, maintained, operated and decommissioned; (b) ‘available techniques’ means those developed on a scale which allows implementation in the relevant industrial sector, under economically and technically viable conditions, taking into consideration the costs and advantages, whether or not the techniques are used or produced inside the Member State in question, as long as they are reasonably accessible to the operator; (c) ‘best’ means most effective in achieving a high general level of protection of the environment as a whole.’

The national implementation of Directive 2010/75/EUinto German law (BImSchG), uses the notions ‘state of the art’ (‘Stand der Technik’
\footnote{§ 3 sec. 6 Gesetz zum Schutz vor schädlichen Umwelteinwirkungen durch Luftverunreinigungen, Geräusche, Erschütterungen und ähnliche Vorgänge (,BIsmSchG'; Federal Immission Control Act): ‘State of the art as used herein shall mean the state of development of advanced processes, facilities or modes of operation which is deemed to indicate the practical suitability of a particular technique for restricting emission levels. When determining the state of the art, special consideration shall be given to comparable processes, facilities or modes of operation that have been successfully proven in practical operation’. Translation provided by Inter Nationes, available at \url{https://germanlawarchive.iuscomp.org/?p=315.}}) 

and ‘best available techniques’ (‘beste verfügbare Techniken’\footnote{§ 3 sec. 6d BImschG}) suggesting that they are not identical but closely connected. Legal scholars agree in that context that state of the art at least corresponds to best available techniques.\footnote{Schulte/Michalk in: Giesberts/Reinhardt, BeckOK Umweltrecht (57th ed. 2020), § 3 BImSchG marginal no. 98; see also Deutscher Bundestag, BT-Drs. 14/4599, 126; 17/8125, 3.}  So the minimum basis for state of the art is the best available technique. 

Similarly, § 3 sec. 28 of the German Act Reorganising the Law on Closed Cycle Management and Waste (KrWG) refers to state of the art as ‘the level of development of advanced processes, installations or modes of operation that gives a reliable indication of the practical suitability of a measure for … in the interest of achieving a generally high level of overall environmental protection’ and enlists criteria for determining the state of the art in Annex 3.\footnote{Translation taken from\url{https://www.bmu.de/fileadmin/Daten\_BMU/Download\_PDF/Abfallwirtschaft/kreislaufwirtschaftsgesetz\_en\_bf.pdf}}  Both, KrWG and BImschG, require  that when applying the criteria listed in their respective annexes, the proportionality between the expense and benefits of potential measures as well as the principle of care and prevention needs to be taken into account. Hence, state of the art also considers the individual means available for the operator.

In order to specify the notion further, it is worth delving further into national interpretations since EU law lacks a binding definition.

Following up the example of German law, legal scholars employ  a three step theory, where  ‘state of the art’ is located between the ‘generally accepted rules of technology (‘allgemein anerkannte Regeln der Technik’) and the ‘state of science and technology’ (‘Stand der Wissenschaft und Technik’).\footnote{Jarass in: Jarass, BimschG (13th ed. 2020), § 3 marginal no. 115.}  ‘Generally accepted rules of technology’ require that a certain technology has stood the test of practice and is generally accepted amongst the majority of experts, however, it does not have to be the best technology available.\footnote{Jarass in: Jarass, BimschG (13th ed. 2020), § 3 marginal no. 115.}  This notion derives from building law and the notion of generally accepted rules of architecture describing the dominating opinion of technical experts. There is a (rebuttable) presumption that technical standards such as DIN-norms amount to generally accepted rules.\footnote{Seibel, Abgrenzung der “allgemein anerkannten Regeln der Technik“ vom "Stand der Technik“, NJW 2013, 3000, 3001.}   In contrast, ‘state of science and technology’ relates to a very high level of protection, that requires to take into account the latest scientific knowledge regardless of whether it is technically feasible and available.\footnote{Cf. Jarass in: Jarass, BimschG (13th ed. 2020), § 3 marginal no. 115; Bundesverfassungsgericht, NJW 1979, 359; Martini in: Paal/Pauly, DS-GVO BDSG (3rd ed. 2021), Art. 25 DS-GVO marginal no. 39d.} 

Placing ‘state of the art’ in between these two notions at normative level confirms that state of the art corresponds at least to the best technique available to the operator in question. 

What renders the determination of ‘state of the art’ somehow ‘tricky’ is its dynamic function.

In the context of the GDPR, it has been argued that data controllers are required to adapt their privacy measures regularly to advances in technology.\footnote{Martini in: Paal/Pauly, DS-GVO BDSG (3rd ed. 2021), Art. 25 DS-GVO marginal no. 39d.}  The dynamic reference thus has the potential to enhance innovation, when data controllers are required to constantly adapt their protection measures.\footnote{Cf. Von Grafenstein, Co-Regulation and the Competitive Advantage in the GDPR: Data protection certification mechanisms, codes of conduct and the ‘state of the art’ of data protection-by-design; in: González Fuster, van Brakel, de Hert, Research Handbook on Privacy and Data Protection Law: Values, Norms and Global Politics.}  In that regard Art. 25(1) GDPR requires that in the context of data protection by design or default, state of the art must be taken into account ‘both at the time of the determination of the means for processing and at the time of processing itself’. This may make it difficult to demonstrate compliance with Art. 25 GDPR if a data controller solely relies on an acquired certification since the certification body does not have to constantly monitor the evolvement of state of the art and whether the data controller (or processor) keeps track with technological progress.\footnote{Cf. Von Grafenstein, Co-Regulation and the Competitive Advantage in the GDPR: Data protection certification mechanisms, codes of conduct and the ‘state of the art’ of data protection-by-design; in: González Fuster, van Brakel, de Hert, Research Handbook on Privacy and Data Protection Law: Values, Norms and Global Politics.} 

Although the NIS Directive lacks a determination of the time when state of the art must be taken into account, the ratio of the security provisions implies that the OESs or DSPS have to adapt the security measures to ensure a level of security of NIS appropriate to the risks posed. Recital 53 specifies in that context that in order to avoid imposing a disproportionate financial and administrative burden on OESs and DSPs, the requirements should be proportionate to the risk presented by the NIS concerned. This is in line with the ‘state of the art’ requiring economically and technically feasible measures considering the operator and his processing operations in question.

\section{Overview}
Assessing if a given technology is useful in a given context is a reoccurring task in many engineering projects. Here structured \ac{TRA} use a pre-defined  scale assigning a level of readiness (\ac{TRL}) to a \ac{ToA}. Initially this assessment is a snapshot in time, which evaluates the current state of a \ac{ToA} in a given context. However, \ac{TRL} can be defined in such a way that they document what needs to be doe to move a \ac{ToA} to the next level. Hence performed periodically, they can showcases the progression of the \ac{ToA} over time by collecting evidences on successes and failures, and by recording materialised and potential risks and their mitigation.\footnote{Cf. Hirshorn/Jefferies ‘Final report of the nasa technology readiness assessment (tra) study team’.}
Moreover, periodically performed TRAs can be helpful even if a technology itself is not changed. They can be used to reevaluate a \ac{ToA} in a changing environment: observing possible threats (if any) for making decision throughout the integration process in a larger system.
 
As per GAO study in 2016, many government agencies are commonly using \ac{TRL} to define the technology maturity within its development life cycle and some organizations also have been tailored the TRL definitions to meet their product development goals but in generality is evaluated keeping NASA’s one to nine scale as a benchmark.\footnote{United States Government Accountability Office,``Best Practices for Evaluating the Readiness of Technology for Use in Acquisition Programs and Projects" (2016) pp.16 <https://www.gao.gov/assets/gao-16-410g.pdf> accessed 26 July 2021.} NASA defines level one as basic conceptual technology research, moving to breadboard validation in laboratory environment at level four and then where the technology is tested, proven and consolidated into a final product, successfully launched and operated at level nine.\footnote{cf. Hansen, M., Hoepman, J.-H., Jensen, M. and Schiffner, S. [2015a], Re- port on the workshop on assessing the maturity of privacy enhancing tech- nologies, in ‘IFIP International Summer School on Privacy and Identity Management’, Springer, pp. 97–110.}\textsuperscript{,}\footnote{ICEAA Professional Development and Training Workshop [2017], TRL vs Percent Dev Cost, Technical report.
URL:\url{http://iceaaonline.com/ready/wp-content/uploads/2017/07/DA05- PPT-Linick-Technology-Readiness-Level.pdf
}.} 

In GAO’s 1999 report, {\it{Best Practices: Better Management of Technology Can Improve Weapon System Outcomes}}, the US department of defence and commercial technology firms demonstrate cases which show that new technologies, rated with higher \ac{TRL}, can be integrated more often successfully in projects.\footnote{United States Government Accountability Office(n 16 ) 24.} 

Moreover, \ac{TRA} can facilitate the coordination between accessors, i.e. charged with the assessment of new technologies, and their managers throughout the development cycle. In particular they can surfaces the key obstacles for a technology to get ready to be used and the risks it might introduce to the project. Thus \ac{TRA}s can provide information related to cost, schedule estimates, and risk assessments.\footnote{United States Government Accountability Office(n 16 ) 23-33.}

\ac{TRL}s were originally developed and deployed by \ac{NASA} to help evaluate risks and project readiness of new technologies for the complex cross-functional tasks, mainly for the purpose of space missions. NASA defined a 9-point scale (level 9 being the most mature technology readiness level) that for the most part suited the needs of NASA’s projects where technology (at the time mostly not or only simple software) is involved\footnote{John C Mankins, ‘Technology readiness levels’ (white paper, 1995)  1-5 \url{https://aiaa.kavi.com/apps/group\_public/download.php/2212/TRLs\_MankinsPaper\_1995.pdf} accessed 15 July 2021} 
These \ac{TRL}s, when applied to different industries as it is without any modifications, produced little benefits in assessing the risk or the technology readiness, for example a more refined \ac{TRL} was designed for Manufacturing readiness (MRL), System Readiness Levels (SRL).{\footnote{Sean Ross,‘Application of system and integration readiness levels to Department of Defense research and development in journal Defense AR Journal’ (2016) Vol(23) 248-273 \url{https://apps.dtic.mil/sti/pdfs/AD1015756.pdf} accessed 15 July 2021.}\textsuperscript{,}\footnote{Sirous Yasseri and Hamid Bahai, ‘System Readiness Level Estimation of Oil and Gas Production Systems in journal International Journal of Coastal and Offshore Engineering’ (2018) Vol(2) 31-44.}}

\ac{TRA} is a systematic process that measures the appropriate level of technical and risk understanding of a system's or product’s technology, by assigning a score such as a \ac{TRL} to the \ac{ToA}. The assigned score needs to be supported by evidence, which is obtained using expert interviews and collecting objective metrics, such as lines of code, proof of correctness, market penetration or reported issues.\footnote{Marit Hansen and others (n 13) 2-4.}\textsuperscript{,}\footnote{John C Mankins (n 3) 1-5.}

Due to the lack of a widely accepted best practice for \ac{TRA} or even a unified definition of \ac{TRL}, the understanding on how to assess \ac{ToA} differs greatly by individual project/product managers, each with their own maturity assessment analysis. As per the study conducted by NASA within their own HQs, this difference in understanding has led to multiple delays in project with cost/resource overruns.\footnote{John C Mankins (n 3) 1-5.} Moreover, this diversity in definitions and methods hinders the sharing of information about a \ac{ToA} with other projects, thus work will be duplicated even in the same organisation. 

It needs to be noted, the above described diversity, is not due to ignorance, but due to purpose incompatibility. In the next section we will analyse the gaps of existing TRLs for the evaluation of \ac{ToA} in IT, i.e. such \ac{ToA} that are mainly or even solely based on software and have cyber security relevant functionalities. 

\section{Gaps in existing TRLs for software technology risk assessment}

Having identified shortcomings with NASA's TRLs  when used outside their original scope, the Department of Defense (US DoD) and US department of Homeland security have further fine-tuned NASA’s TRLs and assessment metrics for maturity evaluation of software and IT-services.\footnote{John C Mankins (n 3) 1-5.}\textsuperscript{,}\footnote{Sean Ross (n60).}             
Various TRLs, IRLs (Integration readiness level), and SRLs (System readiness level) have been developed and deployed to meet these purposes. However, they still seem not to address the key software assessment issues, as reported cost/budget and timeline overruns despite the usage of those modified \ac{TRL}s demonstrate, c.f. GAO (US Government Accountability Office) report highlighting further risk to the overall project development.\footnote{United States Government Accountability Office(n 16 ) 6-27.} 

With the above we can observe some incremental improvements over the vanilla NASA \ac{TRL}s; overall, TRLs have proven useful across industries, which in turn resulted in increase popularity with project mangers. However, we also observe that for information technology, incremental improvements might not be sufficient.\footnote{Marit Hansen and others (n 13) 2-4.}\textsuperscript{,}\footnote{John C Mankins (n 3) 1-5.}\textsuperscript{,}\footnote{Steven Hirshorn and Sharon Jefferies (n 1) 6.} 
Further, it is observed that \ac{TRL}s by itself are poor estimators for IT project success. More specifically, given a current and target \ac{TRL}, it is hard to express the effort and resource required to span the gap, i.e. what effort is needed to lift the \ac{ToA} from the current to the target readiness level.\footnote{Marit Hansen and others, (n 13) 2-4.} 

A major pitfall for vanilla \ac{TRL} are none-functional requirements, i.e. properties of IT systems that are necessary for fulfilling their purpose, but are hard to express as a functionality. Cybersecurity and data protection properties are often expressed as none functional requirements, e.g., ``An unauthorised user \emph{should} not be able to access the customer data base". Often these none functional requirements establish a difference in quality of a software. Hence, a lack of such a quality assessments (such as privacy and security), readiness assessments of IT technology might not deliver good predictions of the overall risks deploying a given technology, which renders the exercise useless.\footnote{Marit Hansen and others, (n 13) 2-4.}\textsuperscript{,}\footnote{John C Mankins (n 3) 1-5.} 
Moreover, todays \ac{TRA} relay largely on expert opinions or interviews. In our methodology, we propose. not only to add a quality dimension but also to ask the experts to support their opinion with evidence. In order to facilitate this process we will provide a collection of measurable indicators, which will be used to demonstrate the transition between one level to the next. Such indicators might be the number of publications referencing the \ac{ToA} for an early stage research \ac{ToA} or the number of users, for a product.

 
Another particularity for IT artefacts is their high degree of modularisation. Modules are often repurposed and hence the environment they are used in is changing. In practice in particular the reevaluation of none functional requirements is insufficient. A high profile example is the software imagemagick\footnote{\url{http://www.imagemagick.org/}.}.  Originally, a command line tool developed in 1987 to manipulate bitmap images (Running on a standalone computer; input provided by the user running the software with their access rights.). Today it turned into a library of tools that can be used to implement image manipulations for web services (Running on a server with potentially privileged access rights, processing input by less privileged users, i.e. anyone from the web that is using the service).  This has led and continuous to leads to security issues with the software as the more than 600 CVEs\footnote{\url{https://cve.mitre.org/}.} filed concerting the software at the time of writing.

 Hence, a \ac{ToA} needs to be evaluated in its intended environment; IT focused \ac{TRA} needs to be customised to fit with the  purpose of the software under evaluation including the environment the IT solution is running in. Moreover, current \ac{TRA}s also do not clarify what steps should be taken to address the issues in case software assessment score does not meet the expectations, i.e., what actions can be taken to recover the milestones missed or risk observed. Without providing guidance on those, a mere assessment will prove less favourable acceptance in the software community.

For more than this, we come to the following aims, will target in developing the a new method for software specific \ac{TRA} and a new flavour of \ac{TRL}: 
\begin{itemize}
\item We need to reduce the experts driven subjectivity during assessment. This can be achieved by turning \ac{TRA} into a evidence based process, thus delimiting biases and influences that could shift evaluation decisions basis specific remarks (for example; independent from assessor or assessee feedback or views).
\item We need to develop a guide/handbook detailing which type of evidence is needed to transition the \ac{ToA} from one level to the next. In the best case, this evidence should be readable as requirements list of ``what needs to be done in order to improve a \ac{ToA} to improve" 
\end{itemize}

\paragraph{From readiness to maturity} Considering the existing gaps with NASA's vanilla TRLs today and software industry requirements, we are redefining existing single scalar of 9 level in NASA's \ac{TRL}. Our scale has 2 dimensions: a readiness dimension, that reflects similar to NSAS's scale, how readily usable a \ac{ToA} is and a quality scale, that reflects how well a \ac{ToA} fits into a given context. For better distinction, we call this new 2 dimensional scale \ac{TML}.  

The proposed methodology shall consider the need to build an effective technical readiness and quality assessments to for IT solutions right from early development phases to a more matured state where resource allocations and cost decisions become critical. 

\paragraph{Target of Assessment (ToA) definition} The first step of an evaluation is the definition of the \ac{ToA}. The foremost here is to answer the question: Which technology needs to be evaluated? It needs to be noted, that in real world scenario, this question is often answered rather naturally in the design and early implementation phase, if we consider a traditional software development cycle. However, in particular considering Security by Design principles, this selection might have a huge impact which technology eventually is selected. Our evaluation method needs to be part of this cycle. More specifically, it might be necessary to select several candidate technologies as \ac{ToA} and base the decision of which to use on our \ac{TML}. However, this candidate selection process is not part of our method. What is important here, is to document the purpose of the technology and the intended environment as part of description of the \ac{ToA}. 

The next basic step is to categorise the \ac{ToA} in the following rough dimensions:

\begin{itemize}
\item Software types identification (Dependent / Independent): example are a) newly developed software, (b) internal / existing or reused (c) commercially-off-the-shelf (COTS) software. 
\item Lifecycle/individual maturity of each software (including when multiple softwares are at play). Example - Is it still relevant or outdated or have just arrived in the market. 
\item Security/Privacy aspect of the software (whether this is the primary goal of the software or not). Example - security requirement for the specific project is high or low?
\end{itemize}


\section{ New Methodology Proposed and Modelling} 


The assessment method proposed here is built on the assessment from the literature study and research work on existing TRLs that have been used for software maturity assessment. Similar to penetration testing the assessors need to answer a set of per-defined questions and need to provide supporting evidence for their answers. Novel for SMART is, that maturity is defined as a vector with 2 elements: readiness and quality. The set of questions is thus split in 2 sections: a \emph{readiness  assessment}, and a \emph{quality assessment}, resulting in individual levels for both and combined in a maturity level.

The questionnaire to assess the readiness of a \ac{ToA} aims to assign a 6 level scale. The decision as to which level the \ac{ToA} is assigned to is based on  set of questions that enquires the software attributes around specific parameters. Similar to the reviewed literature above, we define the following 6 levels: \emph{1. Idea, 2. Research, 3. Concept, 4. Trial, 5. Product, and 6. Outdated maturity state.} 

To these readiness levels, four quality characteristics are added: \emph{1. Protection components, 2. Risks parameters, 3. Operational efficiency, and 4. Enhancement opportunities.} The model will require assessor to score both Readiness and Quality levels every time the assessment is done. For each of the readiness level and quality characteristics one of the three possible maturity state is determined: \emph {1. Negative (-) state – readiness level scores under the determined threshold percentage, 2. Positive (+) state – scores above the determined threshold percentage, and 3.  Neutral (0) state – score lands above the negative but below the positive threshold percentage}. Then the score is calculated based on the responses to the pre-defined set of objective questions for each of the readiness and quality levels aimed to fully assess the maturity state of the software product. In a scenario, when a readiness level scores negative maturity state then the SMART model determines the set of actions required to achieve the positive maturity state by the time of next assessment. Those actions call for varying level of design and quality characteristic checks, securing supporting documents and plan for follow-up assessment. For score arriving at neutral maturity state, the assessor would determine whether to hold the maturity at the current level or to move to the next level depending on the criticality and remaining gaps to the next readiness level.  The follow-up review for the negative score will be done to determine whether the gap has been addressed or not will vary by the software type and complexity and how far or close the team /organization is on those fixes. A more detailed approach to such scenario is covered later in the paper.

In case the readiness level/Quality maturity state scores above the determined certain threshold percentage then the software is determined to have fully met the requirements for that level and is deemed as success/on-path to next level goals/milestones. The combination of score from readiness and quality levels for each readiness level state is detailed further in the paper with example calculations. 

\subsection{SMART Readiness and Quality levels:}

For our evaluation method: the Software Maturity Assessment and Readiness Technique (SMART), we utilise a 6-point scale from earlier research, which we discussed above in legnth.\footnote{Marit Hansen and others, (n 13) 2-4.} 
We selected this scale as it simplifies the process overall and results are easier to interpret by the end user, in fact the definitions of the levels is made with interpretability for a wide audience in mind.  In this section both readiness and quality levels are defined in detail below.

\begin{enumerate}
\item { \textbf {Readiness levels: 6-point scale}} \\
\subitem { \textbf {Idea:}} The \ac{ToA} is considered on the idea level if it exists only as white papers, technical reports or is talked about in blog posts or  at conferences. Ideas can have one of 3 quality possible quality levels: 1) “I\textsuperscript{-}", software classification and objective will be pulled up to align on next steps and advising on what gaps need to be fixed, 2) “I\textsuperscript{0}" for the neutral state, basis software classification a decision will be made to move the maturity to next level or to be put on hold and a re-assessment could be recommended, and 3) I\textsuperscript{+} state means the software under assessment fully matures to the next level.

\subitem { \textbf {Research: }}A \ac{ToA} goes beyond the idea level if it has matured into a Research item.  That is, more advanced scientific studies have been conducted, providing on paper evidence for the feasibility. The problem or customer need has been well defined. The research work analyses the security and other quality characteristics. Similar to the Idea level, Research will have one of the 3 possible quality levels: 1. “R\textsuperscript{-}"– a neutral, 2. “R\textsuperscript{0}"– neutral, and 3. “R\textsuperscript{+}" means positive.

\subitem{ \textbf {Concept: }} At this level, all characteristics of the \ac{ToA} have been fully defined and assessed. A pre-alpha testing has been conducted, e.g. by lab experiments using simulations with artificial data or mockups for a ``smoketes". If a \ac{ToA} reached this level, it should be possible to decide if it is functionally suitability. Starting this level, the software product must achieve a quality score of at least neutral  to advance to next levels. 

\subitem { \textbf {Trial: }}A limited real world test of a feature complete implementation of the \ac{ToA} has been executed to demonstrate the performance, compatibility, and functionality. A beta test has been performed successfully, and its results are recorded as artefacts of the design development and to be supplied during readiness level evaluation. Similar to the Concept level, only those software products that have achieved score neutral or positive for both readiness and quality levels are considered for advancement to the next level.

\subitem { \textbf {Product: }}The \ac{ToA} has been released and is used by a large number of customers. The performance and acceptance of the software is available via customer feedback. The overall maturity evaluation of the \ac{ToA} needs be above a defined threshold for both readiness and quality. 

\subitem{ \textbf {Outdated:}} The maturity at this level means that a \ac{ToA} has become outdated either in performance or requires high operational overhead to maintain the quality requirements as intended. \ac{ToA}s become outdate either by advancement in the technology of the \ac{ToA} (newer products that fulfil the requirements better), or by averse change of the environment, e.g., change of use of a \ac{ToA}, change of the computing platform, or new attacks against certain features of the \ac{ToA}. A \ac{ToA} is transitioning to outdated if it receives negative evaluation either for readiness (i.e. implementation is not available for a new environment) or quality level (e.g. new attack).

\item { \textbf {Quality Evalutation.}} 
As discussed before, we consider the overall maturity of a \ac{ToA} to be composed from a readiness and a quality aspect. These dimensions are not orthogonal, that is for a given overall assessment a minimum quality level is required in all aspects. How to assess these aspects is discussed below.
We split quality metrics into four characteristics:  

\subitem{ \textbf { Security: }} This characteristic has 2 components for assessmment: 1. Protection Goal, and 2. Trust assumptions.  The protection goal aspect considers which assets are protect from whom (which type of attacker) according to the well known characteristics such as: confidentiality, integrity, availability, unlinkability, transparency, intervenability. However, protection goals can be achieved under various trust assumptions, i.e., all components technical/functional/legal that needs to be trusted and to which extent. The maturity and performance are measured in terms of the degree of trust in components or agents, the more trust needed in components the lower the score. To arrive at an overall score we select  lowest level. For example, if Protection gets a final score of “-“ and trust assumptions are rated “0” or “+”, the scoring for Security will be the lowest of the two, i.e., P\textsuperscript{-} or P\textsuperscript{0}. For a P\textsuperscript{+} score we will need to have both the sub-metrics to have at least achieved the “+” score.

\subitem{ \textbf {Risk levels:}} To measure the Threat coverage aspects of the \ac{ToA} the evaluation of 2 characteristics is recommended: Side Effects, and Reliability. Side effects measures the undesirable features arising due to the deployment of the \ac{ToA}. For example, removing latency from an element could expose the data theft risk for some \ac{ToA}s. This metric also measures the composability of the \ac{ToA} with other components or systems. Reliability measure the lifecycle of \ac{ToA} up to which it can function as per specifications without failing or running into errors. It is directly linked to the fault tolerance build in the technology or the recoverability of the system back to its normal state, if and when it crashes or fails to operate. Both of these risks characteristics will be evaluated separately similar to Security metrics and a final score will be determined basis the lowest score of the two sub-metrics.

\subitem{ \textbf {Operational Flexibility/ Efficiency: }}The level measures the maturity of the \ac{ToA} with respect to its operational characteristics, such as: Performance Efficiency, Operability, and Maintainability. Performance efficiency measures the performance of the \ac{ToA} in terms of the resources consumed or cost incurred, for example: storage, computational power, bandwidth, speed etc. Operability is the ease or difficulty in understanding the aspects of \ac{ToA}. This feature helps one in integrating \ac{ToA} with larger systems, example; appropriateness, reconcilability, learnability, technical accessibility and compliance. Maintainability is the degree of effectiveness and efficiency with which the product can be modified or adapted to underlying changes in the overall system architecture. It is measured in terms of modularity, reusability, analyzability, changeability, modification stability, and testability. Inline to scoring logic above, the final maturity status would require all the 3 sub-metrics to have reached “+” state or “0” where the final \ac{ToA} categorisation will be applied to break the tie and allocate maturity status.

\subitem{ \textbf {Enhancement opportunities: }}This level assesses scalability characteristics of the \ac{ToA}, i.e., transferability, and scope. Transferability measures the extent and ease with which the \ac{ToA} can be transferred from one hardware, \ac{ToA}, or environment to another. It measures the portability and adaptability. Scope details the different applications where the \ac{ToA} is applied/can be applied to. This help to assess the gap, if any, with the intended  usages. The scoring logic remains the same as above, lowest score determines the overall score.

The overall maturity evaluation process under SMART applies equal weightage on quality characteristics in general, one might consider changing this wights if a given \ac{ToA} is used in a specific scope, but this comes at the cost that evaluations are less versatile. This is contra the aim  behind SMART: to create a process that could assess a wide range of \ac{ToA}s.

 \item { \textbf {Overall Functioning of the SMART}}.   
 
 Figure \ref{fig: SMART} provides an overview of all SMART stakeholders and the actions that might take place among them. The key idea is, that SMART assessments are used to communicate among these stakeholders. Below, we give a usage example, specifically focused on an active guided development process.
 
SMART can be used as pure assessment tool, using a given set of questionnaires applying them to a \ac{ToA}. This might happen during the design phase of a project. In the best case this initial evaluation finds a perfectly suitable \ac{ToA} for the given use case of high quality and high readiness. Then the evaluation result can be used to e.g. communicate with the competent authorities to demonstrate compliance. However, often developers will identify gaps in functionality, quality or readiness. Here, the strength of SMART is its design to provide guidance during the further integration and implementation process. 

Given an initial evaluation that identified gaps for a \ac{ToA} w.r.t. its intended use, we propose SMART as a guidance tool.
To put SMART at work for a given \ac{ToA} during the development life cycle, developers and assessors will work closely together. Developers will provide status updates on \ac{ToA}, while assessors will refine SMART questionnaires with industry experts/researchers as per ToA under development and continuous evaluation. Evaluation result by assessors are shared with developers for further enhancements. 



 \begin{figure} 

\includegraphics [scale =0.4]{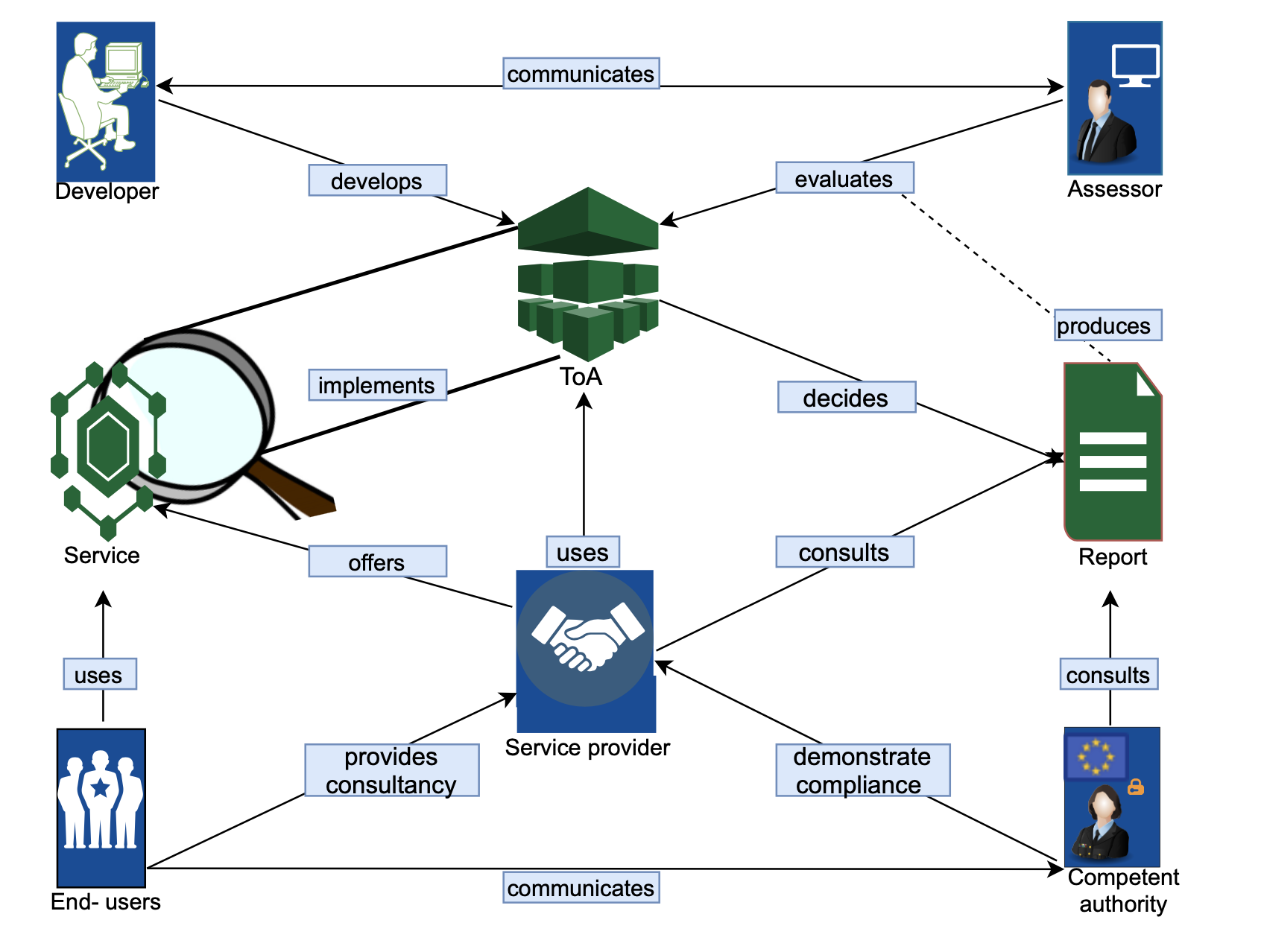}
\caption { SMART}
\label{fig: SMART}

\end{figure}

 \item { \textbf {Scoring/ Evaluation method}} 
 
In order to evaluate the readiness and Quality level of the \ac{ToA}, assessors are asked a set of objective questions that assesses the ToA holistically on all of the readiness characteristics. Each of these questions are framed with binary responses in mind to avoid falling into grey areas. For the responses, assessors are also required to submit proofs/evidences. Evidence can reach from anecdotes, that is single supporting data points to a certain score up to meta studies.
Based on these questionaries  readiness and quality levels are calculated and composed to an overall maturity score for the \ac{ToA}. Currently SMART applies equal weight to each of the questions. However, it is subject to further research to assign more expressive weights. We expect that there is no one size fits all weighting exists and that scoring wights will depend on the application domain. However this might undermine versatility of SMART. However, the highly structured process allows to reuse most of the evaluation (if public) even if wights are changed due to application domain needs. 


Once all the responses are recorded in the questionnaire then model evaluates both readiness and quality maturity score for each of the 6-point readiness levels. The scores are multi-dimensional and provides maturity on both design completion and security/privacy aspects in a single score. Until Concept readiness level, the model does not apply hard constraint on quality assessment factors to be at positive or neutral state, meaning the four quality characteristics could be even in negative maturity state for the readiness level to advance to next level. Beyond concept to Trial readiness level, all the four quality characteristics need to have at least reached neutral maturity state along with the readiness score must be above the pre determined threshold percentage to be  eligible to be promoted to next level. For \ac{ToA} to be moved to Product readiness level, it is a must for all quality characteristics to at least reach a positive maturity state – a mandatory hard constraint built in the model. The reason for these varying quality checkpoints is to ensure that any gaps from \ac{ToA} privacy /security perspective are addressed before design has reached too much advance stage where corrections/updates could result too costly or time consuming.

\end{enumerate}

\section{Conclusion}

We presented SMART, a maturity assessment method to evaluate the technology readiness of IT services their components. SMART will support the communication among stakeholders in the product development cycle to improve overall product quality while reducing budget and timing risks during the development of new services based on existing modules. 

\section{Acknowledgments}

This work has been supported by the Luxembourg National Research Fund as part of EnCaViBS, grant number C18/IS/12639666/EnCaViBS/Cole.










\end{document}